\documentclass[aps,twocolumn,preprintnumbers,amsmath,amssymb]{revtex4}
\usepackage{graphicx}% Include figure files
\usepackage{dcolumn}% Align table columns on decimal point
\usepackage{bm}% bold math
\usepackage{multirow}
\usepackage{color}
\usepackage{ulem}
\usepackage{amsmath}
\usepackage{gensymb}
\begin{document}
\title{Well separated trion and neutral excitons on superacid treated MoS$_2$ monolayers}

\author{Fabian Cadiz$^{1}$}
\email{cadiz@insa-toulouse.fr}
\author{Simon Tricard$^{1}$}
\author{Maxime Gay$^{1}$}
\author{Delphine Lagarde$^{1}$}
\author{Gang Wang$^{1}$}
\author{Cedric Robert$^{1}$}
\author{Pierre Renucci $^{1}$}
\author{Bernhard Urbaszek$^{1}$}
\author{Xavier Marie$^{1}$}
\affiliation{%
%$^1$ Institute of Photonics and Quantum Sciences, SUPA, Heriot-Watt University, Edinburgh EH14 4AS, United Kingdom}
%\affiliation{%
$^1$ Universit\'e de Toulouse, INSA-CNRS-UPS, LPCNO, 135 Av. Rangueil, 31077 Toulouse, France}

%\listfiles
%\nofiles
%\keywords{WSe2, valley dynamics, time resolved photoluminescence, transition metal dichalcogenides, two dimensional materials}
%\date{\today}

\begin{abstract}
Developments in optoelectronics and spin-optronics based on transition metal dichalcogenide monolayers (MLs) need materials with efficient optical emission and well-defined transition energies. In as-exfoliated MoS$_2$ MLs the photoluminescence (PL) spectra even at low temperature consists typically of broad, overlapping contributions from neutral, charged excitons (trions) and localized states. Here we show that in superacid treated MoS$_2$ MLs the PL intensity increases by up to $60$ times at room temperature. The neutral and charged exciton transitions are spectrally well separated in PL and reflectivity at $T=4$ K, with linewidth for the neutral exciton of $15$ meV, but with similar intensities compared to the ones in as-exfoliated MLs at the same temperature. Time resolved experiments uncover picoseconds recombination dynamics analyzed separately for charged and neutral exciton emission. Using the chiral interband selection rules, we demonstrate optically induced valley polarization for both complexes and valley coherence for only the neutral exciton.
\end{abstract}

%\pacs{78.60.Lc,78.66.Li} %Use showkeys class option if keyword

\maketitle
\textit{Introduction.---}
Transition metal dichalcogenide (TMD) monolayers (ML) such as MoS$_2$, MoSe$_2$, WS$_2$ and WSe$_2$ are a new class of two-dimensional semiconductors with a direct bandgap in the visible region of the spectrum \cite{Mak:2010a,Splendiani:2010a,Ross:2013a} and very unique properties. Strong spin orbit coupling combined with a crystal lattice that has no inversion symmetry allows for optical manipulation of the spin and valley degree of freedom in these materials \cite{Xiao:2012a}. In addition to their potential for unconventional, atomically thin and flexible electronics or valleytronics, they also are ideal candidates for optoelectronic and spin-optronic applications. For example, solar cells \cite{Tsai:2014a},photodetectors \cite{Lopez:2013a} and laser  prototypes \cite{Salehzadeh:2015a} based on mono or few-layer MoS$_2$ have been demonstrated.\\

In these ideal 2D systems the optical properties are dominated by excitons, as the electron-hole Coulomb interaction is strongly enhanced due to quantum confinement and reduced screening  \cite{Chernikov:2014a, Hill:2015a}. In addition to neutral excitons (X), also charged excitons (T) contribute to the emission as ML samples are intentionally or non-intentionally doped \cite{Mak:2013a}. Here the exciton linewidth and defect contribution to the emission is very different from one material to another. For example, at temperatures between $4$ and $100$ K ML MoSe$_2$ shows only sharp X and T emission (FWHM 10 meV) separated by about 30 meV \cite{Ross:2013a}. In contrast, the PL of as-exfoliated or CVD grown ML MoS$_2$ over the same temperature range is characterized by a broad exciton peak (FWHM 60 meV) at an energy between $1.88$ and $1.9$ eV \cite{Lagarde:2014a,Cao:2012a,Kioseoglou:2012a, Zeng:2012a}, which stems from overlapping T and X contributions as well as from broad low energy defects-related emission. \\

Very recently, Amani et al. \cite{Amani:2015a} have shown that the room temperature PL intensity of superacid treated ML MoS$_2$ is two orders of magnitude higher than in as-exfoliated layers. We have used this treatment to cure exfoliated samples and find, in addition to the increased room temperature PL intensity, several important improvements of the optical quality, particularly striking at low temperature. First, the emission at $4$ K is now dominated by sharp, well separated trion and neutral exciton emission, very similar to high quality MoSe$_2$ and WSe$_2$ MLs \cite{Jones:2013a,Wang:2015b,Ross:2013a}. Second, the emission at low energy related to defects is greatly suppressed. It has been found, in addition, that the PL intensity of neutral and charged excitons is similar at $4$ K in non treated and treated MLs. Nevertheless, the clear separation of the exciton and trion in treated MLs allows to probe individually valley polarization and dynamics. These experiments on cured ML MoS$_2$ open the door for more demanding optoelectronics and spin manipulation experiments in this material, which were so far impossible to carry out due to the spectrally broad emission stemming from several different contributions.\\

\indent \textit{Samples and Experimental Set-up.---}
Mechanically exfoliated flakes of bulk MoS$_2$ crystals (2D semiconductors) were transfered  onto SiO$_2$/Si substrates using a viscoelastic stamp \cite{Gomez:2014a}. The SiO$_2$ layer ($290$ nm thick) provides an efficient color contrast under white light illumination, allowing an easy identification of MLs. Treated samples were prepared as follows: first, the samples were annealed on a hot plate at $150 \degree$ C during 30 minutes in ambient air.  Then, bis(trifluoro-methane) sulfonimide (TFSI) was dissolved on 1,2-dichloroethane (DCE) at a concentration of $0.2$ mg/mL, and the samples were immersed in this solution at $100 \degree$ C during 10 minutes. Finally, the samples were rinsed in DCE and annealed again during 5 minutes at $100 \degree$ C on a hot plate.\\

A standard micro-PL set-up is used to record the PL spectra in the temperature range $T=4-300$ K. For stationary measurements, MLs were excited with  a $532$ nm cw laser. For polarization-resolved measurements, the flakes were excited by $1.5$ ps pulses generated by a tunable optical parametric oscillator (OPO) synchronously pumped by a mode-locked Ti:sapphire laser. For time-resolved photoluminescence experiments, the flakes are excited by $\sim 1.5$ ps pulses at $430$ nm generated by a tunable mode-locked  frequency-doubled Ti:Sa laser with a repetition rate of $80$ MHz \cite{Lagarde:2014a}. In all cases, the excitation spot diameter is $\leq1\mu$m, i.e considerably smaller than the ML size of typically $\sim 10~\mu$m$\times10~\mu$m, and the PL emission is dispersed in a spectrometer and detected with either a cooled Si-CCD camera or a Hamamatsu C5680 streak camera for time-resolved experiments. \\

\begin{figure*}
\includegraphics[width=0.9\textwidth]{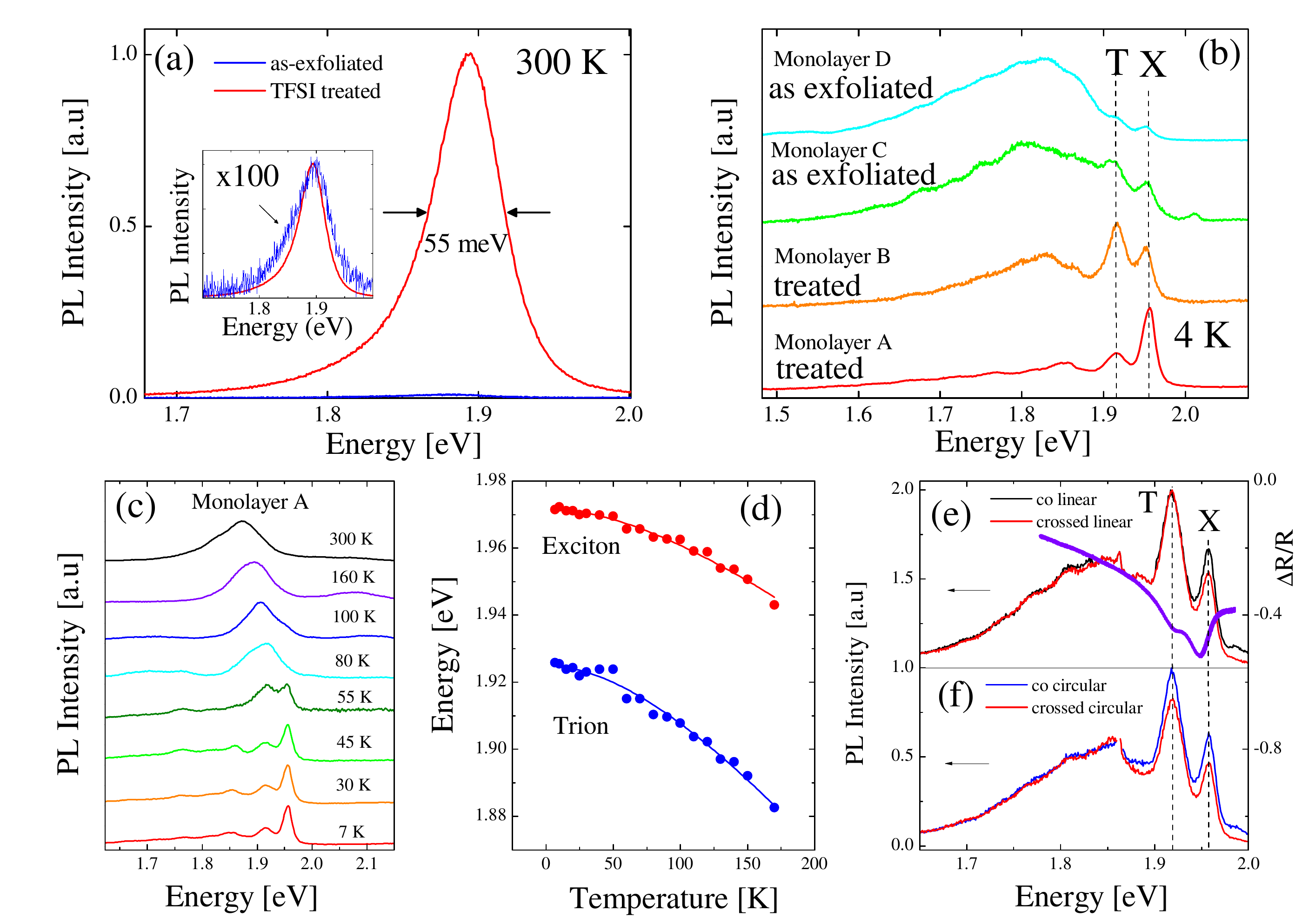}
\caption{\label{fig:fig1} (a) PL spectrum of an as-exfoliated (blue) and superacid treated (red) MoS$_2$ monolayer at room temperature in ambient conditions. For better comparision, the inset shows the same curves in which the emission before treatment has been multiplied by 100, revealing a reduction in linewidth from $\sim 70$ to $\sim 55$ meV after treatment. (b) Typical spectra for as-exfoliated and treated samples at $T=4$ K. Whereas the intensity is maximum at low energies for as-exfoliated samples, exciton and trion emission are the dominant features on treated samples. The curves have been normalized and shifted for clarity. (c)   Normalized photoluminescence of a treated MoS$_2$ monolayer vs. temperature. (d) neutral exciton (red) and trion (blue) peak position vs. temperature with fits (solid lines) using Eq.(\ref{eq1}). (e) Polarization-resolved photoluminescence under linearly polarized excitation at $T=4 K$. Also shown is the reflectivity contrast (violet, solid line), in which the neutral exciton peak is clearly visible and is accompanied by a less intense shoulder at the trion position.  (f) Polarization-resolved photoluminescence under circularly polarized excitation at $T=4$ K.
}
\end{figure*}

\indent \textit{Results and Discussion.---}
Panel (a) of Fig.\ref{fig:fig1} shows the PL spectrum at room temperature for a MoS$_2$ ML before (blue) and after acid treament (red). On 7 different samples, it is found that the spectrally integrated intensity is enhanced by 20 to 60 times on TFSI treated MoS$_2$ MLs for a low  power excitation ($0.1$ $\mu$W) at $532$ nm. This dramatic increase in PL efficiency is accompanied by a reduction of the linewidth (from $\sim 70$ to $\sim 55$ meV), as can be seen in the inset of Fig.\ref{fig:fig1} (a) in which the PL spectrum before treatment has been amplified by two orders of magnitude for better comparison.\\

\noindent
As shown in panel (b) of Fig.\ref{fig:fig1}, the typical spectrum of two as-exfoliated samples (C and D) at a temperature of $4$ K presents a broad low energy emission ascribed to defect-related trapped exciton states, which overlap strongly with exciton and trion emission, preventing a clear identification of these two peaks. In contrast, the spectra of treated samples (A and B) have a significant reduction of the low energy emission, and the neutral exciton ($1.967$ eV) and trion ($1.924$ eV) can be thus nicely resolved, with linewidths of $16$ and $37$ meV, respectively. In all cases, the incident laser power was kept at $10$ $\mu$W. We emphasize that the same PL intensity for X and T is found for treated and as-exfoliated MLs at $T=4$ K. The exciton and trion separation, of $42$ meV, is larger than the $30$ meV found typically on WSe$_2$ and MoSe$_2$, in agreement with previous measurements \cite{Mak:2013a}. This suggests a high non-intentional doping of the monolayer. The clean spectra of the exciton and trion in our acid treated samples cleary represent a dramatic improvement in optical quality.\\

\noindent
The temperature dependence of the PL emission, shown in panel (c) of Fig.\ref{fig:fig1} for a treated sample (A) suggests that both X and T contribute to the spectrum at room temperature, in agreement with the findings of Mak et al.\cite{Mak:2013a}. This provides a possible explanation to the very high sensitivity of MoS$_2$ to dielectric environment, gate voltages and even physisorption of air molecules, which can cause efficient charge tuning on the monolayer and therefore an important modulation of the PL intensity  \cite{Tongay:2013b,Mak:2013a}. This is a major difference with standard MoSe$_2$ and WSe$_2$ MLs in which the T contribution vanishes above $150$ K \cite{Robert:2016a}. The variation of both X and T energy position with temperature, shown in panel (d) of Fig.\ref{fig:fig1} can be well reproduced by the Varshni relation:

\begin{equation}
E_{\mbox{ex(tr)}}(T)= E_{\mbox{ex(tr)}}(0)-\frac{\alpha T^2}{T+\beta}
\label{eq1}
\end{equation}

\noindent
Solid lines in panel (d) of Fig.\ref{fig:fig1} are fits to our data, with the parameters $\alpha=0.39$ meV/K and $\alpha=0.61$ meV/K for X and T, respectively, and $\beta=253$ K, similar to the fitting parameters found by Kioseoglou et al.\cite{Kioseoglou:2012a}, although only one broad peak at the trion energy was observed in that study at all temperatures. \\

Both X and T transitions can also be identified by the differential reflectivity spectrum, defined as $(R_{\mbox{flake}}-R_{\mbox{SiO$_2$}})/R_{\mbox{SiO$_2$}}$ and shown in panel (e) of Fig.\ref{fig:fig1} (violet curve). Indeed, both transitions are related to a high density of states, whereas no signature of transitions at the energy of the localized emission is visible. \\

We now focus on the ability to study the polarization properties of X and T lines separately, which has been elusive so far in MoS$_2$.
In TMD MLs, the circular polarization of an absorbed/emitted photon can be directly associated with selective carrier excitation/population in one of the two non-equivalent K valleys of the Brillouin zone, $K^+$ or $K^{-}$. As shown in Fig.\ref{fig:fig1} (f),
both exciton and trion peaks of a treated monolayer exhibit large circular polarization in PL under circular excitation at $2.101$ eV, while the localized states emission is largely unpolarized, in agreement with previous measurements in as-exfoliated MLs \cite{Lagarde:2014a, Mak:2012a}.
%Note that several studies of valley-selective circular dichroism of MoS$_2$ revealing high circular polarization of the luminescence have been performed with an excitation energy that, according to the present findings, corresponds to a nearly-perfect resonance with the neutral exciton energy \cite{Cao:2012a}, \cite{Lagarde:2014a}.
 Whereas excitation with \textit{circularly} polarized light initialize valley polarization, excitation with \textit{linearly}  polarized light results in exciton alignment, also termed valley coherence \cite{Jones:2013a}. In Fig.\ref{fig:fig1} (e) we show linearly polarized emission of the X following linearly polarized excitation. No linear polarization is observed for the peak we attribute to the trion, as can be expected for this three-particle complex. The linear polarization analysis observed in our cured samples allows us to confirm the attribution of the two sharp peaks to X and T complexes.\\

\begin{figure*}
\includegraphics[width=0.9\textwidth]{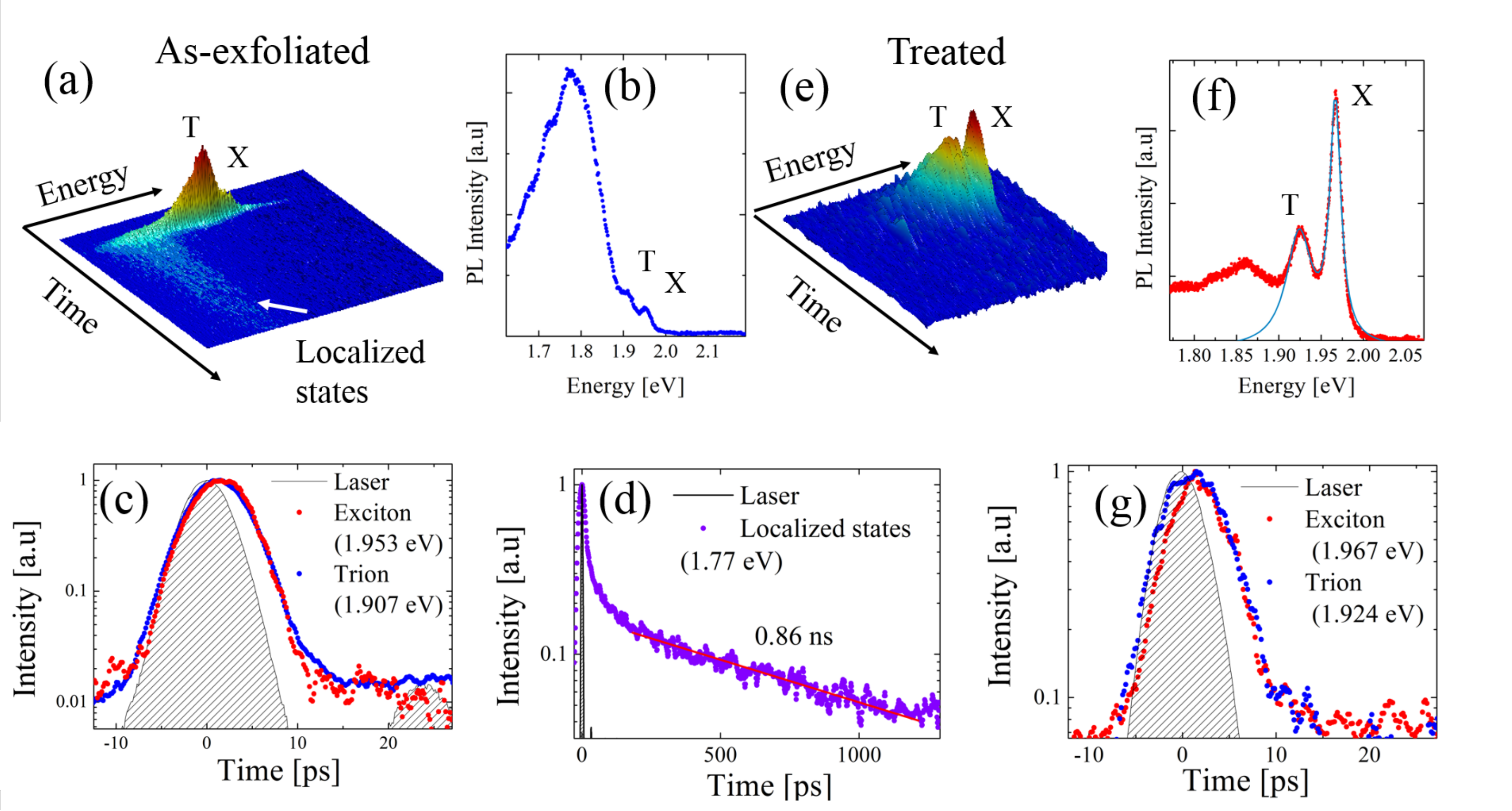}
\caption{\label{fig:fig2}  T=4~K. (a) 3D representation of the time-resolved emission of an as-exfoliated MoS$_2$ monolayer. The long-lived emission at low energies is characteristic of localized excitons on defects (b) time-integrated spectrum of (a). (c) intensity of trion (blue) and exciton (red) lines as a function of time. Also shown is the laser pulse. (d) decay of the localized emission shown in panel (a). (e) same as (a) for a treated monolayer. (f) time-integrated spectrum of (e) (in red). The blue solid line is a two-lorentzian fit. (g) same as (c) for a treated monolayer.
}
\end{figure*}

Time-resolved photoluminescence was performed in order to study the dynamics for X and T complexes separately. The time-averaged laser power was kept in the $\mu$W range in order to avoid sample degradation. Panel (a) of Fig.\ref{fig:fig2} shows a 3D representation of the time resolved emission for an as-exfoliated MoS$_2$ monolayer at $T=4$ K, which presents, in addition to a broad trion peak and a smaller exciton feature, a slowly decaying shoulder at lower energies which corresponds to localized excitons \cite{Lagarde:2014a}. Panel (b) of Fig.\ref{fig:fig2} shows that the time-averaged spectrum is dominated by this localized emission that hinders the identification of trion and exciton peaks, both presenting a decay with a very fast dynamics, as can be inferred from the profiles shown in panel (c) of Fig.\ref{fig:fig2}. The localized emission's temporal evolution, shown in panel (d), has a decay lifetime on the ns scale. When time-resolved measurements are performed on treated MLs, the emission of localized states is highly suppressed, and emission comes mainly from well defined exciton and trion peaks, shown in panels (e) and (f) of Fig.\ref{fig:fig2}. Note that in this particular flake, exciton emission is more intense than trion's, probably caused by an important depletion of the unintentional n-doping. 
Again, a very short emission time of $3.2 \pm 0.3$ ps, almost limited by our temporal resolution, is observed for the exciton and trion (panel (g)), just like in as-exfoliated MoS$_2$ samples. This demonstrates that the low temperature exciton dynamics in treated and non-treated MLs is not dominated by non-radiative recombination on the defects which are suppressed by the acid treatment. This exciton dynamics can be interpeted by a fast radiative decay at low temperatures as a result of the strong exciton oscillator strength in TMDC MLs, which is consistent with recent measurements performed on MoSe$_2$ and WSe$_2$ MLs\cite{Robert:2016a}, and theoretical predictions \cite{Wang:2016b,Palummo:2015a}.


\indent \textit{Perspectives.---}
In conclusion, MoS$_2$ monolayers have been treated with TFSI superacid. In addition to the enhancement of the PL intensity by nearly two orders of magnitude at room temperature, it has been shown  that there is a significant reduction of localized states emission at low temperatures, allowing to obtain well resolved emission for both exciton and trion complexes. This should open the door for a better understanding of the exciton physics on ML MoS$_2$, an ideal candidate for atomically thin opto-electronic devices.\\
\indent \textit{Acknowledgements.---} 
We thank ANR MoS2ValleyControl and ERC Grant No. 306719 for financial support. X.M. also acknowledges the Institut Universitaire de France. F.C and P. R thank the grant NEXT 
No ANR-10-LABX-0037 in the framework of the « Programme des Investissements d'Avenir".

\bibliography{2dmaterials}% Produces the bibliography via BibTeX.
\end{document}